\documentclass{article}
\usepackage{graphicx} % Required for inserting images
\usepackage{natbib}

\title{What Do Black Holes Teach Us About Wigner's Friend?} 
 \author{Emily Adlam \thanks{Philosophy Department and Institute for Quantum Studies, Chapman University, Orange, CA92866, USA \texttt{eadlam90@gmail.com} }}
\date{\today}
\usepackage{amsmath} 
 
\date{\today} 
\setcitestyle{round}
\usepackage{mathrsfs}

\begin{document}

\maketitle

\begin{abstract}

Recently, Hausmann and Renner have pointed out that several famous paradoxes relating to black holes have a similar character to various Extended Wigner's Friend paradoxes. In this paper I consider what the connection between these things could teach us about the Wigner's Friend scenarios. I argue that  if we take the analogy between these cases seriously, the black hole paradoxes appear to favour a certain class of response to the Wigner's Friend scenario - specifically, those which posit intrinsic relationality, rather than effective and emergent relationality, and also those which posit some kind of retrocausality.

\end{abstract}

\section{Introduction}

Recently, \cite{hausmann2025firewallparadoxwignersfriend} have pointed out that several famous paradoxes relating to black holes have a similar character to various Extended Wigner's Friend paradoxes. Given these connections, studying the black hole paradoxes offers an interesting way to get some new insight into the Wigner's Friend paradoxes. 

In particular, a central feature of both types of paradox is that certain pairs of measurement outcomes cannot ever be simultaneously witnessed by the same observer, and this feature is essential to most putative  solutions of the paradoxes. But the \emph{reasons} why these outcomes cannot be known by the same observer seem quite different in the black hole case as compared to the Wigner's Friend case. So if we believe that the basic principles of these situations are similar, comparing them may tell us something about the nature of the physical mechanisms involved in the problem, and thus it can potentially help inform us about the right response.

I will begin in section \ref{Wigner} by introducing the Wigner's friend paradoxes and arguing that if we insist on the first-person universality of quantum mechanics, these paradoxes tell us that quantum states must sometimes be relativized at least in an effective empirical sense. Then in section \ref{black} I will introduce the black hole paradoxes referenced by \cite{hausmann2025firewallparadoxwignersfriend} and compare them to the Wigner's Friend scenarios. I will argue that if we take the analogy between these cases seriously, the black hole paradoxes appear to favour a certain class of responses to the Wigner's Friend scenario - specifically, those which posit \emph{intrinsic} relationality, rather than effective and emergent relationality, and also those which posit some kind of retrocausality.

\section{Preliminaries}

The goal of this paper is to understand what we can potentially learn from the analogy drawn by Hausman and Renner. Thus throughout this paper, I will be making two key assumptions: 

\begin{itemize} 

\item In their presentation  of the black hole paradoxes, Hausman and Renner  make a number of claims about what certain observers would be expected to  see in certain circumstances, and these claims depend on specific choices about how to describe black hole physics. In this article I will largely assume that these claims are correct.

\item I assume that, if the analogy set out by Hausmann and Renner holds, then it is plausible that both kinds of paradox will have similar solutions. 

\end{itemize}

It must be emphasized that both of these assumptions are quite substantive and could certainly turn out to be false. Regarding the first assumption, at present most black hole physics is not  observationally accessible, and thus all assertions about what would be observed near and within  black holes are necessarily speculative. Moreover, the presentation of Hausmann and Renner relies upon a certain paradigm for black hole physics. First, it is consistently assumed that evolution is everywhere unitary - so, for example, spontanteous collapse approaches are ruled out by fiat. Second, the black hole is described in a  `global' manner, which relies on the invocation of an event horizon  whose location at a given time depends on happenings in the far future and distant in space. While this paradigm is common, it is not uncontroversial \citep{pittphilsci13195}, and thus of course one possible way to address the black hole paradoxes as presented by Hausmann and Renner is simply to deny some of their assumptions about black hole physics.

However, it is precisely these assumptions that allow Hausman and Renner to set up an operationalized scenario for black holes which looks similar to the Wigner's Friend case, so if the assumptions are wrong there is simply no analogy to be drawn. Moreover, if the resolution to the paradoxes ultimately hinges on specific details of black hole physics, it seems unlikely that a comparison with the Wigner's Friend scenarios will yield much insight into that resolution. Thus since my goal  in this article is to consider  what might be learned from an analogy between black holes and the Wigner's Friend scenarios,  I will largely grant the assumptions about black hole physics which are needed to set up these operationalized scenarios, in order to focus on assessing how strong the analogy  is and what it might tell us. In particular I will largely follow Hausmann and Renner in making use of the global event horizon paradigm, although I will reconsider this in section \ref{teleology}.

Regarding the second assumption, of course we cannot know for sure that two similar paradoxes must have similar resolutions - even if we grant that there is a strong resemblance between them, it is always possible that the resemblance is a mere coincidence. Still, there are some good reasons to seek a similar resolution. First of all,  Hausmann and Renner are following a well-established methodological path by establishing an operational resemblance between two different scenarios - for each pairing of a black hole paradox and a Wigner's friend paradox, they identify a map between observers and then exhibit a set of quantum operations available in both cases which appear to lead to a contradiction in both cases. And  it is a very common idea in physics that  phenomena which stand in such a relation of operational similarity should, where possible, be explained in similar ways. \cite{spekkens2019ontological} argues that a  principle of roughly this kind, which he dubs the `ontological identity of empirical indiscernibles,' shows up in the work of \cite{leibniz2000leibniz} and played a significant role in the formulation of special relativity \citep{Einstein1905}. Moreover, the idea that operational similarity should, where possible, guide and constrain ontological descriptions is also  common in the modern field of quantum foundations \citep{2015Entrp..17.7752P,Catani_2023,schmid2020unscrambling}. So given that  this is a standard  principle of reasoning in the field already, it seems reasonable to adopt a similar principle here as well and see where it leads. 

Moreover, in addition to this operational similarity, there is also a substantive \emph{conceptual} similarity. For each pair of paradoxes, it not merely the case that sets of similar operations lead to a contradiction with quantum theory, but also the contradiction is a contradiction \emph{for the same reason} - in the first pair of paradoxes, because a system is measured in two incompatible bases, and in the second pair of paradoxes, because an observer can perfectly predict the outcomes of measurements in two incompatible bases. This indicates that these scenarios are similar not only at the level of operationally observed effects, but also in terms of the way in which they are naturally represented within the theoretical structure of quantum mechanics, which gives us additional reason to think there is likely to be some connection between their solutions.

There is also a further conceptual similarity which seems particularly relevant here:  in both the black hole case and the Wigner's Friend case, we avoid empirical contradictions in these scenarios precisely in virtue of the fact that there are pairs of measurement outcomes which cannot ever be accessed by any individual observer. Given that this feature is so essential to maintaining \emph{empirical} consistency, it is very natural to think that it will also be relevant to maintaining consistency at a conceptual level, so it will very likely play a central role in the resolution of the paradox on both the black hole side and the Wigner's Friend side. And evidently if indeed the same feature is central to the resolution of the paradox in both types of scenario, then there will be similarities between those resolutions. So  here we have not only an operational similarity, but  also strong reasons to expect that the solutions are likely to be conceptually similar.

Of course, there are also some reasons for caution. In particular, the black hole paradoxes involve gravitational degrees of freedoms whereas the Wigner's Friend paradoxes do not appear to do so, and one might think that moving to a whole new regime of physics would undermine any attempt to draw connections between them. However, in some ways the fact that the black hole paradoxes involve gravitational degrees of freedom makes the emergence of similar phenomenology in the two cases even more striking – one might not have expected much commonality between these regimes, so if some commonalities emerge nonetheless, that is a strong signal that there is some relevant connection. Perhaps the similarity arises because in fact gravitational physics \emph{is} relevant to the Wigner's Friend paradoxes in some non-obvious way, or perhaps it arises because gravity itself is ultimately quantum and these paradoxes are linked to fundamentally quantum effects common to all quantum theories, but either way, the connections across different regimes seem striking enough to merit further investigation. 

Thus, while fully acknowledging that the assumptions listed above could be wrong, I think it is nonetheless a worthwhile project to see what follows if we take seriously the idea that there is a connection between these two kinds of paradoxes. One reason for this is that both the Wigner's Friend paradoxes and the black hole paradoxes both pertain to regimes that we cannot currently probe experimentally, and thus in both cases the discussion around them is quite unconstrained; looking for solutions which can apply in a similar way across the two contexts is therefore valuable as a way of imposing some structure on this discussion. Moreover, since the black hole paradoxes are central to a number of ongoing debates within quantum gravity research programmes, and meanwhile the Wigner's Friend paradoxes are an important recent development in the ongoing debate over the measurement problem, this seems like a promising way to get a handle on possible connections between quantum gravity and the measurement problem, thus hopefully paving the way for the two fields to inform one another in useful ways. 

 I also emphasize that in this paper my focus is to try to  understand what we can learn from  the putative analogy between black holes and Wigner's Friend, not to offer an overall assessment of the merit of different approaches, and thus I will largely disregard advantages and disadvantages which are not directly related to this analogy. In particular, it should of course be acknowledged that some kinds of responses to these paradoxes are much easier to implement than others. For example, in general, Everettian approaches  will always have the notable  advantage that they are ready `out of the box' - that is, although we might have conceptual worries about the Everett approach, we probably don't have to do much new work at the technical level to apply it to the scenarios discussed here. By contrast, nearly every other possible way of resolving these paradoxes will require alterations to physics, and potentially very significant ones which we are otherwise incentivized to avoid. These considerations are important and should certainly form a part of the overall decision we make about which routes to pursue, but I will generally put them aside here in order to focus on the putative analogy which is the subject of this paper.

\section{Wigner's Friend Paradoxes \label{Wigner}}

The original Wigner's Friend paradox involves the following scenario: an observer, Alice, performs a measurement  $\{ |0 \rangle \langle 0 |, |1 \rangle \langle 1  | \}$ on a two-dimensional system $S$ originally prepared in the state $a | 0 \rangle + b |1 \rangle$, with both $a$ and $b$ non-zero. This is a projective measurement and thus standard quantum mechanics suggests the system is now in the state given by the standard state update rule for projective measurements, conditioning on the outcome of the measurement \citep{Nielsen} -  that is, if Alice sees the outcome $0$ then the subsequent state of system $S$ is $| 0 \rangle$, and if Alice sees the outcome $1$ then the subsequent state of system $S$ is  $| 1 \rangle$. 

However, if Bob observes from the outside and describes this whole interaction using unitary quantum mechanics,  he will predict that $S$ and Alice are in the following entangled superposition state: 

\[ \psi_{SA} = a | 0 \rangle_S | 0 \rangle_A + b |1 \rangle_S | 1 \rangle_A \]

Here, $|0\rangle_A$ and $|1 \rangle_A$ are states of Alice corresponding to `seeing outcome 0' and `seeing outcome 1' respectively. This scenario was historically  regarded as puzzling or paradoxical, since Alice presumably believes that  a single definite outcome to her measurement has occurred, and yet $\psi_{SA}$ seems to represent Alice and S as being in a state where no definite outcome has occurred.

One obvious way to resolve this paradox would be to simply  deny that quantum mechanics is universally correct - for example, we could posit a wavefunction collapse or something of that kind, in which case $\psi_{SA}$ is not  the correct state to predict Bob's measurements on $S$ and Alice. However, it is not clear that the black hole case offers any new insight into this kind of solution, so in this article I will focus on the approaches which seek to preserve the universality of quantum mechanics. More specifically, I will focus on approaches which maintain what I will call the `first-person universality' of quantum mechanics: that is, they ensure that quantum mechanics \emph{as applied by any individual observer} will always make the right predictions \emph{for measurements performed by that observer}. 

In order to maintain first-person universality, we must say that in the Wigner's Friend scenario, either the state  $| 0 \rangle$ or  $| 1 \rangle$  gives the right predictions for the result of any subsequent measurement that Alice makes on $S$, but also the state $\psi_{SA}$ gives the right predictions for any subsequent measurement that \emph{Bob} makes on $S$, even though these states are incompatible. Thus if we want to maintain first-person universality we have no choice but to accept that at least at some effective empirical level, quantum states are sometimes relativized, in the following specific sense: the quantum state that correctly predicts the results of one observer's  measurements on a system $S$ at a given time $t$  is not always the same as the quantum state that correctly predicts the results of another observer's measurements on that same system $S$ at the very same time $t$. 

For example, relativization of this kind appears in both the Everett interpretation and the de Broglie-Bohm model. In the Everett interpretation, the states we use to predict the outcomes of measurements performed by a certain observer must be relativized to the branch of the wavefunction the observer is in, so the state which correctly predicts the results of measurements performed by an observer in one branch is not usually the same as the state which correctly predicts the results of measurements performed by an observer in another branch. Similarly in the de Broglie-Bohm approach, while we can always make the correct predictions using the full quantum state and all the positions of the Bohmian particles, if we want to simply write down the quantum state which best describes the predictions for measurements made by a given observer without specifying the full distribution of the  particles, we will need to use `conditional wavefunctions' relativized to the particles composing the observer in order to arrive at effective wavefunction collapses in certain scenarios  \citep{D_rr_1992, articlenorsen}.

Note that this does not mean that either of these views is relational in any deep or metaphysical sense: the point is merely that they posit certain emergent phenomena for which it is sometimes convenient to use a `relational' description, even though there is nonetheless an underlying absolute description. So the Everett and de Broglie-Bohm approaches are relational only a fairly weak, high-level way - in particular,  there are two distinct axes along which an approach might be `more relational' than these approaches.  First of all, Everett and de Broglie-Bohm posit that quantum states are relational only at an \emph{effective} level: they maintain that there always exists an underlying observer-independent quantum state, while allowing that are some circumstances in which it is appropriate to use different effective quantum states to predict outcomes for  different observers. By contrast,  views such as Relational Quantum Mechanics (RQM) posit that quantum states are \emph{inherently} relational, meaning that quantum states are by their very nature descriptions of a system's dynamics relative to an observer, and therefore it simply does not make sense to postulate an underlying absolute quantum state  as Everett and de Broglie-Bohm  do. 

Second, Everett and de Broglie-Bohm posit relationality which is only \emph{dynamical}, meaning that although some quantum states and hence some dynamics are relativized, the actual underlying facts of reality are not relativized. By contrast, some authors responding to the Wigner's Friend scenario have posited a stronger form of \emph{fact-based} relationality according to which  facts themselves - indeed, perhaps all facts -  must be relativized to an observer or reference system \citep{Cavalcanti_2021,brukner2015quantum,Bene2001-BENAPV,kochen1985new,1996cr}.

 Thus we have a schema of possible types of relational views \citep{adlam2025kindrelationalitydoesquantum}; I will shortly consider whether the black hole paradoxes might help us choose between these approaches:

\begin{center}
\begin{tabular}{ c |  c  | c }
  & \textbf{Effective  } & \textbf{Inherent  } \\ \hline
   \textbf{Dynamical}  & Everett interpretation  & RQM+CPL\footnotemark  \\ 
&   de Broglie-Bohm &   \\ \hline
\textbf{Fact-based}  & X  & Orthodox RQM  \\ 
& & Perspectival modal interpretations \\
& & Neo-Copenhagen/perspectival approaches
\\ 
\end{tabular}
\end{center}  
 \footnotetext{This refers to the version of relational quantum mechanics suggested by \cite{Adlam-2023}, which includes an additional postulate of Cross-Perspective Links as discussed further in section \ref{2WF}.}

 \subsection{Extended Wigner's Friend Paradoxes}

Recently a number of Extended Wigner's Friend paradoxes have been proposed, aiming to sharpen the reasoning above with no-go theorems demonstrating that certain clusters of assumptions are incompatible with the predictions of quantum mechanics. 

It is helpful to separate the EWF paradoxes into two classes, based on the type of assumption they make: 

\begin{itemize}

\item \textbf{Third-person universality:}  the derivation of the `paradox' relies on the assumption that quantum mechanics produces correct joint predictions for \emph{any} pair of measurement outcomes.\footnote{\cite{schmid2023review} refer to what I call `Third-Person Universality' as `Born Inaccessible
 Correlations' - i.e. the requirement that when two measurements are made in parallel,
the two outcomes always arise with a joint frequency given
by the Born rule, even if no single observer could access
both outcomes, even in principle.}

This class includes the original discussion of \cite{Wigner1995}, the theorem of \cite{Deutsch1985QuantumTA}, the theorem of \cite{2018qtcc}, and the theorems of  \cite{leegwater2018greenberger}  and  \cite{ormrod2022nogo}.

\item \textbf{First-person universality:}   the derivation of the `paradox' relies only on the assumption that quantum mechanics as applied by each individual user of the theory always produces correct predictions for measurements performed by that user.

This class includes the Local Friendliness theorem of  \cite{Bong_2020} and the extension by \cite{ying2023relating}.

\end{itemize}

Note that Wigner's Friend paradoxes based on third-person universality can be resolved immediately by positing that quantum mechanics is universal only in a first-person sense, as described in section \ref{Wigner}, and not necessarily in a third-person sense. One straightforward way to do this is to posit inherent relationality, since it is an immediate consequence of such a view that quantum mechanics can only be universal in a first-person and not a third-person sense.  However, Wigner's Friend paradoxes based on first-person universality of course cannot be resolved in this way. 

Both of the Wigner's Friend paradoxes described by \cite{hausmann2025firewallparadoxwignersfriend} belong to the class of paradoxes based on third-person universality, and thus they can be resolved in this straightforward way.  However, I will argue shortly that the same is not true for the two black hole paradoxes described by \cite{hausmann2025firewallparadoxwignersfriend}, and thus it is likely that they will require a more complex solution. 

\subsection{Paradox 1-WF}

Paradox 1-WF, which is based on a theorem of  \cite{Deutsch1985QuantumTA}, involves a situation in which Alice performs a measurement on a quantum system $S$ in the diagonal basis, and then Bob reverses Alice's measurement procedure before performing a measurement on $S$ in the computational basis. Thus it seems that Alice and Bob have been able to measure a single system in two incompatible bases, which is impossible according to standard quantum mechanics. For example, imagine that we are writing down  a quantum state for the system as a whole after this procedure:  as explained by  \cite{hausmann2025firewallparadoxwignersfriend}, there is no quantum state we can possibly assign which obeys the standard quantum-mechanical measurement update rule conditional on both Alice's observed outcome and also Bob's observed outcome. 

However, although there is certainly a conceptual tension present here, there is no outright contradiction - that is, no individual observer in this situation will ever see something which directly contradicts the predictions of quantum mechanics as applied by that observer. This is because  no individual observer can possibly know both Alice and Bob's outcomes at once, since the way in which Bob's measurement is constructed requires that he erases Alice's outcome before obtaining his own. This means that we get a contradiction with quantum mechanics only if we assume third-person universality, so we can resolve the paradox by adopting a relational approach which allows us to reject third-person universality. 

Moreover, note that in this scenario  Alice and Bob's outcomes do not ever overlap at a single time, and thus either effective or inherent relationality will serve to resolve the paradox. For example, the Everett interpretation (an effective-dynamical approach) resolves the paradox by saying that Alice's post-measurement state is relativized to the branch that she is in; two different branches exist simultaneously for the two possible measurement outcomes, both ceasing to exist when Bob reverses Alice. So although we do have a global quantum state which must ultimately predict everything, nonetheless there is no time at which a single well-defined outcome of Alice and a single well-defined outcome of Bob all appear as part of any one global quantum state, and thus quantum mechanics as understood within the Everettian picture does not make any joint predictions across Alice and Bob's outcomes. Meanwhile, Relational Quantum Mechanics (which is naturally understood as either an inherent-dynamical or an inherent-fact approach)  resolves the paradox by saying that Alice's outcome is part of her relative state and Bob's outcome is part of his relative state and there is never any individual relative state describing both outcomes at once, so again quantum mechanics does not make any joint predictions across Alice and Bob's outcomes. Either way, the solution is to deny that the ordinary quantum predictions must apply across measurements performed by different observers at different times when the outcomes of those measurements can never be compared.  
  
 \subsection{Paradox 2-WF \label{2WF}}

Paradox 2-WF, which is similar to the theorem of \cite{2018qtcc}, involves a situation in which observers Alice and Charly perform computational basis measurements on a pair of entangled qubits $Q_A$ and $Q_C$ inside their respective closed laboratories. If third-person universality holds, Alice's outcome will be correlated with Charly's outcome as predicted by quantum mechanics. Bob then reverses Alice's measurement and performs a diagonal basis measurement on the qubit $Q_A$. If third-person universality holds, Bob's outcome will be correlated with Charly's outcome on $Q_C$ as predicted by quantum mechanics. On runs when Bob obtains a certain outcome $O$ to his measurement, Bob can then infer with certainty what Charly's outcome was, and subsequently he can also infer with certainty what Alice's outcome was. Finally we have Darwin reverse Charly's outcome and prepare to measure the qubit $Q_C$. If third-person universality holds then Darwin's outcome will be correlated with Alice's outcome as predicted by quantum mechanics. Thus if third-person universality holds, then on runs when Bob obtains the outcome $O$ he is able to use this chain of correlations to predict with certainty the outcome of a computational basis measurement performed by Darwin on $Q_C$. 

But Bob can also simply apply quantum mechanics directly to all of the other observers and qubits together, and this will allow him to predict with certainty the outcome of a diagonal basis measurement by Darwin on $Q_C$. And diagonal and computational basis measurements are incompatible, so it looks as though Bob can predict with certainty the outcomes of two incompatible measurements on the same qubit, which is impossible according to standard quantum mechanics. 

Note that in order to perform this chain of reasoning, Bob is making an assumption which \cite{hausmann2025firewallparadoxwignersfriend} call `Consistency of Knowledge': if agent 1 knows that agent 2 has correctly arrived at a quantum description of a system $S$ and this description predicts that a certain measurement $M$ will have an outcome $O$, then agent 1 should also predict that the measurement $M$ will have an outcome $O$. For example, this is what Bob uses when he infers Charly's outcome based on his own outcome. `Consistency of Knowledge' is closely related to what I have called `third-person universality,' except that it puts the emphasis on what observers believe and predict rather than on the correlations themselves. 

As in the previous case, outright contradiction is avoided here - that is, no individual observer in this situation will see something which directly contradicts the predictions of quantum mechanics as applied by that observer. This is because no individual observer can possibly know both Alice and Bob's outcomes at once, since the way in which Bob's measurement is constructed requires that he erases Alice's outcome before obtaining his own. Similarly, no individual observer can know both Charly and Darwin's outcomes at once. This means that we get a contradiction with quantum mechanics only if we assume third-person universality, so we can resolve the paradox by adopting a relational approach which allows us to reject third-person universality and thus also  `Consistency of Knowledge.'  For if we say that the state that Bob writes down based on the result of his measurement of $Q_B$ is relativized to him, then it can only be used to predict the outcomes of \emph{Bob}'s subsequent measurements, not Charly's measurements, so the inference from Bob's outcome to Charly's outcome is illegitimate.  Bob may perhaps be able to predict the outcomes of Charly's measurements if he has enough information to determine the correct quantum state relative to Charly, but he cannot expect that the quantum states relative to \emph{Bob} will generally make the correct predictions for Charly's measurements; similar points apply to the other steps in the chain of inferences, so the reasoning is not valid. 

 In light of this, I think the name `Consistency of Knowledge' should be treated with care, for there is nothing inconsistent about denying `Consistency of Knowledge' once we posit that a quantum description used by one observer predicts the outcomes of measurements performed \emph{by that observer} and not necessarily anyone else.  In particular, \cite{hausmann2025firewallparadoxwignersfriend} suggest that if Consistency of Knowledge fails, then it would not be possible for `experimentalists (to) gain data and communicate them to their theorist colleagues' or for `the theorists (to) communicate their conclusions based on this data back to the experimentalists.' However, this is not necessarily the case. If Consistency of Knowledge fails, what that means first and foremost is that the correct predictions for the outcomes of my measurements on a given system are not necessarily the same as the correct predictions for the outcomes of your measurements on a given system. This does not imply that we cannot communicate - communication is still possible as long as there exists some kind of `measurement' I can make on you (e.g. asking you a question) such that the result of that measurement reliably matches the value of some variable relevant to you, such as a result that you have seen during a previous observation.  

To see this, consider the example of RQM. This is a form of inherent-dynamical or inherent-fact relationalism, so it denies third-person universality and hence Consistency of Knowledge is violated. But as  discussed in \cite{Adlam-2023}, it is possible to add to RQM an extra postulate called `cross-perspective links' (CPL) which ensures that agents can communicate with one another in appropriate physical interactions. So this version of RQM does violate Consistency of Knowledge, but it is explicitly the case that observers are still able to communicate data and conclusions with each other.

One might object at this juncture that in order to do science, it is not enough that we should be able to communicate measurement results: we must be able to communicate our \emph{reasoning} and rely on conclusions drawn by other observers. This seems right; but note that violations of Consistency of Knowledge do not necessarily undermine it. In the case described above, Bob can communicate the reasoning that he uses to predict the outcome of his own measurement, and Alice can understand and agree with his reasoning, appreciating that he has made the right prediction for his own measurement while also understanding that this might not necessarily be the right prediction for her own measurements. Similarly, in effective-dynamical  relational approaches such as the Everett or Bohmian picture, we know that decoherence will naturally produce a shared quasi-classical regime within which observers can communicate both their results and their reasoning about those results; the fact that Consistency of Knowledge fails in some special cases outside of this quasi-classical regime does not undermine the ordinary processes of sharing of information and reasoning which take place within this regime. Thus failures of Consistency of Knowledge are not in and of themselves threats to the possibility of intersubjective collaboration. 
  
\section{Black Hole Paradoxes \label{black}}

 \cite{hausmann2025firewallparadoxwignersfriend} compare each of the Wigner's Friend paradoxes described above to an analogous black hole paradox. Here I use labels to indicate the correspondences suggested by  \cite{hausmann2025firewallparadoxwignersfriend}:

\subsection{Paradox 1-BH \label{1BH}}

 Paradox 1-BH is a version of an experiment proposed by \cite{Hayden_2007}. In this scenario,  Alice carries a qubit $Q$ into a black hole which is older than the Page time, and once inside she performs a measurement in the computational basis on it. Meanwhile,  on the outside Bob collects all of the Hawking radiation that the black hole ever produces. Within the paradigm for black hole physics assumed by Hausmann and Renner, if Bob collects enough radiation he will eventually  be able to reconstruct the qubit $Q$, and he can then measures it in the diagonal basis. So now the qubit $Q$ has been measured in two incompatible bases, which is impossible according to standard quantum mechanics\footnote{Another way of putting the paradox is to note that it looks as though by this method Bob has succeeded in creating a perfect copy of Alice's qubit $Q$, which should be impossible given the quantum no-cloning theorem \citep{Scarani_2005}.}.

This paradox is similar in form to Paradox 1-WF, since that case also involves measuring the same qubit in two incompatible bases. And once again, outright contradiction is avoided because no individual observer can possibly know both Alice and Bob's outcomes at once. Thus this scenario violates the predictions of quantum mechanics only if if we assume third-person universality, so we can resolve the supposed paradox by adopting a relational approach which allows us to reject third-person universality. 

This is, in effect, the approach that is taken in the `complementarity' approach to black hole physics, which  is based on the idea that the exterior and interior descriptions of the black hole are actually consistent because they cannot be directly compared to each other. In fact, the word `complementarity' in the literature refers to a cluster of related ideas - as characterized by \cite{MuthukrishnanForthcoming-MUTUBH}, there are at least two quite distinct notions here. First, \emph{operational} complementarity involves maintaining that there is no contradiction as long as `no violations of the accepted principles of quantum physics can be detected by any observer,  whether outside or inside the black hole' \citep{Hayden_2007}. This operational view simply declines to admit as meaningful any statements about relationships between  observations which cannot ever be compared by any single observer. Second, \emph{descriptive} complementarity involves maintaining that `the exterior and infalling descriptions are descriptively consistent (as opposed to just operationally consistent) with each other and with QM' \citep{MuthukrishnanForthcoming-MUTUBH}.  There are different ways in which this descriptive version might be cashed out, but it often seems to be associated with quite a metaphysically radical picture. For example, inspired by holography, some presentations of complementarity suggest that the exterior and infalling descriptions of the black hole are really just identical descriptions of exactly the same physics \citep{Lowe_2016,Susskind_1994} - that is, the exterior and the interior of the black hole are in some sense to be identified with each other, so there  simply could not exist any joint, perspective-neutral description of the inside and the outside of the black hole considered as distinct systems. What the operational and descriptive notions of complementarity seem to have in common is the idea that it is simply not meaningful to formulate a joint description relating what is seen inside the black hole to what is seen outside the black hole. 

So let us consider how these forms of complementarity are related to solutions that have been posed in the Wigner's Friend case.  First,  although technically a retreat to operationalism could resolve the problem in the Wigner's Friend  case in much the same way as it does in the black hole case, this option has not featured prominently in discussions of the Wigner's Friend  paradoxes, perhaps for sociological reasons. On the other hand, the `descriptive' approaches to complementarity seem closely related to various forms of `fact-based' relationalism which have been proposed in the Wigner's Friend  case, since fact-based relationalism is similarly predicated on the idea that it is not meaningful to formulate a joint description relating pairs of measurements in the case where they can't be directly compared.

But here the analogy to the Wigner's Friend case is  instructive, because we saw in that case that there is another option: we could adopt a form of \emph{inherent-dynamical} relationalism, where we maintain that it \emph{is} meaningful to formulate a joint description relating this pair of measurements, but we deny that it is appropriate for that description to be formulated within the standard quantum formalism, because there is nothing that such a description could be relativized to. We can resolve the paradox without denying the possibility of joint descriptions altogether, because the mere existence of a joint description cannot violate the predictions of quantum mechanics if we simply deny that it is appropriate to use quantum mechanics to formulate this description in the first place. This way of thinking about the paradox certainly has something in common with complementarity, but it looks a little different from either the operational approach or the typical formulation of the descriptive approach, and thus in this article I will take it that `complementarity' should be thought of as corresponding to only one subclass of possible relational resolutions to these paradoxes, in order to allow that there might be other types of relational resolutions with metaphysical commitments different from those typically associated with complementarity.

If we now look in more detail at the analogy between 1-BH and 1-WF, there are also some key differences between these paradoxes: in particular, although it is essential in both cases that no single observer can ever compare the outcomes of Alice and Bob, the  mechanisms which prevent joint access of the outcomes are quite different in these two cases. In paradox 1-WF, the reason why no observer can access both measurement outcomes is because Bob erases the information about Alice's outcome when he performs his measurement on her. So one might naturally think that something significant is happening to the physical system itself during this physical process of erasure - as \cite{Hadamardsa} puts it `It’s hard to think when someone Hadamards your brain.' For example, this is exactly what happens in the Bohmian account of this scenario: the  result of the first measurement is recorded in the position of the corresponding Bohmian particle, but when Bob reverses Alice's measurement the particle is moved away from that position so Alice's outcome is not subsequently accessible to any future measurement.

By contrast, in paradox 1-BH the reason why no observer can access both measurement outcomes is because one outcome is obtained inside the black hole and the other outcome is obtained outside the black hole, and conventional wisdom says that information can never escape from a black hole\footnote{ There are some approaches \citep{Ashtekar_2025,Ashtekar_2018} which push back against this idea and suggest that information can in fact escape, in the hope of preserving unitarity. But Paradox 1-BH could be regarded as an argument against such approaches, since the problem seems significantly worse if it is in fact possible to get the result of Alice's measurement out of the black hole so it can be compared with Bob's.}. Note that one might be worried that Bob could just wait until he has reconstructed $Q$ and then enter the black hole, at which point he could tell Alice about his measurement outcome, and thus it would be possible for a single observer to observe both outcomes after all - \cite{Hayden_2007} argue that this depends quite sensitively on the timing of various processes in and around the black hole, but it does not seem to be clearly ruled out. In section \ref{2BH} I will discuss a similar case in which the outer observer does enter the black hole, but for now let us simply suppose that  Bob cannot ever compare notes with Alice. 

Furthermore, note that in paradox 1-BH Bob does not perform his measurement by acting directly on Alice or on the black hole containing her; he performs all his operations on the radiation that has escaped from the black hole, a system which is completely distinct from Alice and the physical qubit that she measured. So we cannot simply say that there is a variable of Alice representing her outcome which  gets scrambled by a literal physical interaction with Bob, since Bob does not interact with Alice at all. Moreover, the paradigm for black hole physics employed by Hausmann and Renner says that nothing of particular  significance happens locally to Alice when she passes the event horizon\footnote{A few dissenting approaches \citep{Almheiri_2013,bousso2025firewallsgeneralcovariance}  posit a `firewall' at the horizon, and in this kind of approach we can perhaps say that the information gets physically scrambled, which makes Paradox 1-BH  seem more closely analogous to Paradox 1-WF.} - and indeed  `locating the event horizon by any combination of strictly local measurements is impossible in principle, no matter how ingeniously the instruments are arranged and precisely the measurements made' \citep{sep-spacetime-singularities}. So we also cannot explain what happens  by positing that the information about Alice's outcome is physically disturbed; some other type of explanation will be needed.  

One way of pushing back against this conclusion would be to adopt a version of descriptive complementarity which says that  the interior and exterior of the black hole are literally identical \citep{Susskind,AMPS}. In that case, perhaps we can in fact say that Alice's outcome gets scrambled by a literal physical interaction with Bob, since Alice herself is identical with the radiation outside of the black hole on which Bob acts\footnote{This surely  raises some quite puzzling questions about how to understand consciousness in connection with this kind of strong complementarity thesis, but let's suppose  that there is a way to make sense of Alice's presence in this situation.}. Still, it should be noted that this version of descriptive complementarity itself involves what looks like quite a strong form of non-locality, for it suggests that by acting on the outside of the black hole Bob can  produce a change to the interior of the black hole, with no locally mediated process passing from the exterior to the interior to carry the influence. Arguably this form of non-locality is less problematic than other kinds, because there is no way of constructing a global foliation to specify which events inside the black hole are simultaneous with events outside of the black hole \citep{Mathur_2009}, and so  it is not strictly speaking true to say that Bob can \emph{instantaneously} produce a change to the interior of the black hole\footnote{Thanks to an anonymous reviewer for pressing this point.} -  but nonetheless it does seem to be the case that something quite radically non-local is in action. So it remains true that in order to understand what is happening here it is necessary to posit some kind of global, perhaps somewhat atemporal effect which is rather different from the simplistic idea of the brain as an individual physical system which is disturbed by a Hadamard gate being performed on it.

Moreover, it should be kept in mind that the `event horizon' cannot be defined in an instantaneous way: typically, the region inside the event horizon is defined as `the whole of the spacetime manifold, minus the causal past of future null infinity' \citep{sep-spacetime-singularities, pittphilsci13195}. That is, the interior of the black hole includes all and only those points such that you cannot send a signal from them to some point in the infinite future, thus formalizing  the heuristic idea that a black hole is an area of spacetime from which it is impossible to escape. Evidently this way of defining the event horizon means that its location depends on  facts about future history, and thus, for example, `where the event horizon horizon is located today depends on what I will throw in the black hole tomorrow' \citep{sep-spacetime-singularities,pittphilsci13195}. So in paradox 1-BH, the explanation for why no observer can ever access both measurement outcomes draws on global and teleological facts about the structure of spacetime, and apparently cannot be explained by any local description of the qubits involved.  Therefore if we suppose that these two similar cases 1-WF and 1-BH should be explained in similar ways, the comparison suggests that in paradox 1-WF there also need not be any literal physical scrambling of the system's variables; what matters is simply the  \emph{teleological} fact  that nobody can ever access these two variables together. That is, paradox 1-WF by itself does not force us to adopt a teleological explanation since the `local scrambling' story is always available, but if we want to explain it in the same way as paradox 1-BH then we do have some motivation to look for a retrocausal or teleological account.

Now, at this juncture one might worry that there is a circularity here: since it is necessary in order to set up the BH paradoxes to rely on the teleological notion of an event horizon, it is not so surprising that as a result of analogizing the WF paradoxes to these BH paradoxes we conclude that some kind of retrocausal or teological phenomenon is indicated. In effect the question is being begged by presupposing teleology in the setup\footnote{Thanks to an anonymous reviewer for raising this point}. 
  
As a first response, it should be noted that ultimately the paradoxes BH-1 and BH-2 are \emph{operational} in character: the event horizon is not used directly to draw metaphysical conclusions but rather to arrive at conclusions about what would be learned by various observers under various operationally-characterized circumstances. Moreover, in general we only need to appeal to facts about the future if we want to be able to pin down the \emph{exact} location of the event horizon, since there are various quasi-local methods that can be used to arrive at an approximate location \citep{Nielsen_2009,Bengtsson_2011}. And it is plausible that the operational predictions used in the formulation of the paradoxes ought to be fairly robust against sufficiently small variations in the location of the event horizon, since the protocol does not seem to require that anything is done exactly on the event horizon; so it may remain open to say that the use of the global paradigm is only convenient rather than strictly necessary to set up the paradox. 
  
Additionally, although it is true that  in general there is no straightforward way to replace the globally-defined event horizon with a local description, we can do so if we limit ourselves to a special class of symmetric spacetimes \citep{pittphilsci13195}. And since paradox 1-BH is a thought experiment in any case, we can surely choose to use that kind of black hole. It  doesn't matter that there are no actual black holes of this kind in our world, because the point of the paradox is to probe the internal consistency of black hole physics, and  if the theory is consistent then it should surely be able to give consistent descriptions of worlds more symmetric than ours. So we can in principle simply choose to focus on an example where local descriptions are available in order to avoid any appeal to teleology in describing the paradox. 

Thus there may be ways to set up this paradox without presupposing that the physics is necessarily teleological in nature. And yet  the argument given above still plausibly applies to a paradox set up in this way, for even if we don't have to appeal to teleological considerations to determine the approximate or exact position of the event horizon, it remains the case that the \emph{reason} why this horizon has the significance it does is still bound up with the teleological fact that nothing inside that horizon can escape and so no observers can ever compare the results of the measurements. That is, while of course black hole physics remains speculative and one way or another the argument must depend on unverifiable predictions about what observers would learn in this scenario, there do seem to be possible options for setting up the paradox and arguing for a retrocausal resolution which do not simply beg the question. 

Finally, I would argue that if in fact it turns out that we do have to invoke some fundamentally teleological physics in order to set up the analogy between the black hole case and the Wigner's Friend case, that in is itself is a very a relevant datum: it is a striking observation that \emph{if} we make certain assumptions about black hole physics, we arrive at some phenomenology that looks very similar to the phenomenology expected in a very different context, and this is, it itself, a good reason to take those assumptions seriously. In particular, it is notable that although the retrocausal response to the Wigner's Friend paradoxes has always been  on the table, it has received comparatively little attention compared to other possible responses, so the fact that we see this connection between Wigner's Friend and black hole scenarios if we posit some kind of retrocausal effect may help motivate further enquiry into retrocausal and teleological approaches.

Another reason to think the retrocausality connection deserves further attention comes from the fact that there already exists a prominent teleological approach to the black hole paradoxes, known as  the `final state proposal' \citep{Horowitz_2004,Bousso_2013}. This is based on the idea of adding to ordinary quantum mechanics a final state, specified in this case only for the interior of the black hole, and representing the infalling system and the fields inside the black hole as maximally entangled; this state has the effect of teleporting the state of the matter system to the Hawking radiation outside of the black hole (the mechanism is similar to quantum teleportation, except that no final operation needs to be performed and thus no classical information needs to be sent).  It could therefore be interesting to see if something similar to the final state proposal could furnish a resolution to various different Wigner's Friend scenarios. 
 
What is more, on the quantum foundations side there exists a similar teleological way of resolving the Wigner's Friend paradoxes - the `Lorentzian solution to the quantum reality problem,' a putative solution to the measurement problem  put forward by \cite{Kent}. Kent's approach posits that the course of history is selected according to the outcome of a `measurement' performed on the quantum state at the end of time. This means that events which are not recorded in the state at the end of time fail to be realised as part of the actual course of history. So in Paradox 1-WF, since Alice's measurement is later erased, Kent's approach says that the measurement does not  occur at all and thus has no outcome. And in Paradox 1-BH, since Alice's measurement takes place inside a black hole and all evidence of it when the hole evaporates, Kent's approach also says that the measurement does not really occur at all and thus has no outcome. Now, it is evident that Kent's proposal is different from the final state proposal because Kent's approach seems to predicts that there is nothing much going on inside a black hole at all, whereas the final state proposal adds extra structure inside the black hole in the form of a final state. But nonetheless there are interesting commonalities between these approaches, and thus once again, these connections may strengthen the motivation for looking more closely at retrocausal and teleological solutions to the Wigner's Friend paradoxes, since approaches such as Kent's are still comparatively underexplored.

\subsection{Paradox 2-BH \label{2BH}}
 
Paradox 2-BH is intended as an operationalized version of the firewall paradox \citep{AMPS}, which is an argument  originally put forward with the intention of strengthening previous black hole paradoxes. Almheiri et al argue that complementarity is not enough to resolve all the conceptual problems with black holes, because it a paradox can be derived even  within the perspective of just one observer. It should be noted that Hausmann and Renner's presentation is not the only possible operationalization of the firewall paradox, and we will encounter some alternatives shortly. 

 Paradox 2-BH as initially described by Hausmann and Renner actually uses two  observers. Alice goes into the black hole and distills a qubit $Q_A$ which is maximally entangled with a given qubit $Q_R$ outside of the black hole. Meanwhile, outside of the black hole Bob has collected all of the Hawking radiation it has ever produced, and within the paradigm for black hole physics assumed by Hausmann and Renner, if Bob collects enough radiation he will eventually  be able to construct a qubit $Q_B$ which is also maximally entangled with $Q_R$. Bob measures $Q_B$ in the diagonal basis and sends a record $R$ of his outcome into the black hole. Meanwhile Alice measures her qubit in the computational basis and waits until she can perform a measurement on Bob's record $R$. Now Alice can use the results of her two measurements on $R$ and $Q_A$ to predict with certainty the results of two incompatible measurements on the qubit $Q_R$, which is impossible according to standard quantum mechanics. 

As noted, the prediction that $Q_B$ is maximally entangled with $Q_R$ depends on the paradigm for black physics assumed by Hausmann and Renner. Two assumptions in particular are important:

 \begin{itemize}

\item \textbf{Unitarity}: The evolution of a black hole and all fields from the creation to the asymptotic future is a Haar-typical unitary. 

\item \textbf{EFT:} The fields outside the stretched horizon are described by effective field theory.

 \end{itemize}

 These are relatively standard assumptions in the study of black holes, but as we shall shortly see, there are some open questions about exactly how they should be applied in the context of the black hole paradoxes. 

 Paradox 2-BH is similar in form to Paradox 2-WF, since in that case  we also end up in a scenario where it appears that Alice can predict perfectly the results of two incompatible measurements on a single qubit. Again, outright contradiction is avoided because no individual observer can ever possibly know the outcomes of the measurements on $Q_A$, $R$ and $Q_B$ all at once, so again it seems that the paradox can perhaps be resolved by adopting a relational approach which allows us to reject third-person universality. However, the relational solution is a little more complex in this case. I will now argue that if we suppose  these two similar cases 2-WF and 2-BH should be explained in similar ways, we can draw two important conclusions: first, this comparison again supports some kind of teleological or retrocausal account of these paradoxes, and second this comparison supports inherent relationality over effective relationality. 

\subsubsection{Teleological/Retrocausal Account \label{teleology}}

  \cite{hausmann2025firewallparadoxwignersfriend} argue that  paradoxes 2-WF and 2-BH are similar because   `Consistency of Knowledge' plays a significant role in both of them. In paradox 2-BH, Alice arrives at a prediction for the result of a measurement on $Q_R$ in the diagonal basis by first using her measurement on Bob's record $R$ to infer what Bob's outcome was, and then inferring what prediction he would make for the result of a measurement on $Q_R$, and then taking on this prediction as her own.  So one might think we can resolve this paradox by adopting a relational approach in which Alice's state predicts the outcome of a computational basis measurement on $Q_R$ performed by Alice herself, and Bob's state as inferred from $R$ predicts the results of a diagonal basis measurement on $Q_R$ performed by Bob, but neither state predicts with certainty the results of two incompatible measurements on $Q_R$ which can be performed by a single person, and thus there is no contradiction with quantum mechanics as applied by any individual person.

However, as noted above the firewall paradox was originally intended to demonstrate a paradox within the perspective of just one observer, and indeed Hausmann and Renner later acknowledge that there also exist possible operationalizations of the paradox using just one observer \citep{bousso2025firewallsgeneralcovariance}. By contrast,  paradox 2-WF certainly could not be formulated using only one observer, and this should perhaps clue us in to the fact that `Consistency of Knowledge' cannot really be playing the same role in both paradoxes. That is, in paradox 2-WF, the reason we must appeal to another observer is because we must make reference to specific definite outcomes that are actually obtained by Alice and Charly. The paradox cannot possibly be made to work without making claims about what Alice and Charly see, because in this scenario there is no physical mechanism for communicating the results of Alice and Charly to Bob, and therefore  we cannot rely on Bob simply collapsing the wavefunction for himself. Thus a relational solution may succeed here because we can simply deny that the observations made by Alice and Charly have to be related in any particular way to the observations made by Bob and Darwin, since no individual observer will ever be able to check that relation. 

By contrast, in the black hole case the physical system $R$ is actually sent to Alice, so in principle if we find the right way of describing unitary evolution relative to Alice then she can simply collapse the wavefunction herself and arrive at the contradiction. Thus it is not so obvious that we have  to appeal to definite outcomes obtained by other people in this case. After all, Alice has all the same information as Bob about the nature of the black hole, the procedure that Bob is going to perform and so on, so it  might seem that Alice can simply write down a quantum description of Bob, $Q_R$, $Q_B$, $R$ and the environment all relative to her, and then evolve it forward to arrive at the prediction that \emph{relative to herself} the record $R$ should end up maximally entangled with qubit $Q_R$. That is, one might think there is some reasonable way of applying quantum mechanics from Alice's point of view which leads to the conclusion that it is true  even relative to Alice that $R$ and $Q_B$ are maximally entangled, in which case Consistency of Knowledge need not be invoked at all, since the relevant predictions can all be made within the perspective of a single observer.

In fact there are complexities here because the Unitarity assumption above requires us to make reference to the asymptotic future outside of the black hole, and since Alice is inside the black hole then the asymptotic future outside of the black hole is of course inaccessible to her, so one might think there is something suspicious about having her use Unitarity to arrive at the conclusion that $Q_B$ and $R$ are maximally entangled (relative to her). However, this conclusion seems to depend quite sensitively on difficult and still unresolved questions about how we should stitch together various different parts of spacetime to arrive at a single unitary description relative to a given observer. After all, if we are allowed to use the asymptotic future to calculate a state relative to Bob which then determines a measurement outcome that can be communicated to Alice, then there is evidently some sense in which the asymptotic future is in fact causally connectible to Alice and thus one might think we should be allowed to use it in writing down a unitary description of $Q_R$, $Q_B$, $R$ and  the environment relative to her. Indeed, if we do not allow Alice to use these assumptions in writing down her relative description then it's not clear \emph{what} we can say about the state of $R$ relative to her - if we simply maintain that Alice does not have access to the exterior of the black hole and thus can't make any predictions about it, then it would seem that Alice has no way to predict anything at all about the state of $R$ when it comes in to her from the outside. And if we give up on assigning a state to $R$ it seems as though we have already given up on the first-person universality of quantum mechanics, in which case the paradox is already dissolved simply in virtue of the failure of quantum mechanics itself to make predictions.  

And in any case, if we are really worried about the status of the Unitarity assumption as employed by Alice we can always just use the single-observer operationalization of the paradox instead. For example, \cite{bousso2025firewallsgeneralcovariance}  gets rid of Bob entirely and has all the measurements performed by Alice alone; she purifies the Hawking radiation and collects the qubit $Q_B$ outside of the black hole, then she takes $Q_B$ with her into the black hole, and then she collects $Q_R$ and $Q_A$ and carries out the protocol as before.   Finally she measures $Q_A$ and $Q_B$, and using these two outcomes it seems that she can predict with certainty the outcome of her own subsequent measurement on the qubit $Q_R$ in the computational basis and also in the diagonal basis. In this version of the paradox  we have arrived at a violation of the first-person universality of quantum mechanics without  invoking a second observer at all\footnote{Perhaps one might try to undermine this by positing some kind of disunity between different temporal parts of Alice such that Alice's memory of her own prior measurement does not have to be veridical. However, at this point it would seem we have so little stability that it would seem virtually impossible to make sense of scientific practice, so I think it is hard to take this seriously as a viable response.}.

Thus although 2-WF is indeed a paradox about the extent to which we can make inferences about what other observers have observed, 2-BH does not really seem to be the same kind of paradox: even though Hausmann and Renner formulate it with two observers, it is, like the original firewall argument, ultimately a problem about an inconsistency arising within the perspective of a single observer, or alternatively  a problem concerning a breakdown in unitary quantum mechanics where it simply fails to make predictions in certain kinds of cases. This means that certain kinds of solutions that work for Paradox 2-WF will not work in the same way for Paradox 2-BH - for example in Paradox 2-WF we can fully resolve the paradox by adopting some form of relationalism about quantum states, whereas in Paradox 2-BH simply relativizing quantum states to observers does not seem adequate by itself, since this  will not fix problems arising within the perspective of a single observer. This is in some ways not very surprising, since the original purpose of the firewall argument was precisely to argue that complementarity is not adequate to resolve all the black hole paradoxes, but since we  saw above that the class of relational approaches may include more than just complementarity, it is worth further emphasizing that it seems unlikely that any other form of relationalism about quantum states would suffice here either.

With that said, \cite{hausmann2025firewallparadoxwignersfriend} suggest that if we relativize \emph{more} than just quantum states there may be a possible way to make the relational solution work even in the single-observer case. That is, suppose we  stipulate that not only the quantum state but also the assumptions of Unitarity and EFT used to calculate the appropriate quantum state are `relativized,' in the sense that  they hold true only if they are actually verified. That is, Alice may have personally collected the qubit $Q_B$ outside of the black hole where one might think the Unitarity assumption would apply, but if she does not perform appropriate verification measurements then it need  not be true relative to her that the evolution is a Haar-typical unitary, and thus it also need not be true that $Q_R$ and $Q_B$ are maximally entangled relative to her. In particular, this relational approach requires us to say that even if some \emph{other} observer performs a measurement and finds that the evolution is a Haar-typical unitary, this does not imply that it must be so for \emph{Alice}. Moreover, note that if Alice chooses to go into the black hole to retrieve the qubit $Q_A$, then she cannot possibly check that the evolution outside the black hole is a Haar-typical unitary, since she can only monitor this up until the time when she actually enters the black hole. So if $Q_R$ and $Q_B$ are maximally entangled relative to her only when she checks the asymptotic radiation, it follows that it will never be the case that she has access to all three of the qubits $Q_R, Q_A$ and $Q_B$ on an occasion when $Q_R, Q_A$ and $Q_R, Q_B$ are both maximally entangled relative to her, so no contradiction will ever arise. 

However, this proposal faces two significant problems. First, there is an urgent need to offer some general account of how we are supposed to know which features of our theoretical descriptions are supposed to be relativized for the purpose of any given prediction. For example, in the Wigner's Friend scenario described in section \ref{Wigner}, it is assumed that if Alice performs a measurement on $S$, then as a result of that process Bob will find that when he measures Alice and $S$ they are maximally entangled. There is no requirement in this case that Bob should separately perform an additional test to verify that Alice really did measure $S$ - finding that they are maximally entangled \emph{is} the test. So one might similarly think that in the black hole case, if the black hole has evolved in the usual way then as a result of its evolution Alice will find that $Q_B$ and $Q_R$ are maximally entangled - she should not have to do an additional test to verify that the black hole  evolved as expected, since finding that they are maximally entangled \emph{is} the test! Thus there is a need to explain what is the  relevant difference between these two scenarios. 

Perhaps the  difference is that in the Wigner's Friend case Bob  measures the whole Alice-$S$ system, whereas in the black hole case, if we do not adopt the kind of complemtarity approach which literally identifies the interior with the exterior, there is a distinct physical system, i.e. the radiation at asymptotic times, that Alice does not measure at all in the case where she does not check the evolution outside of the black hole. Or perhaps the relevant difference is that we are now invoking a theory beyond ordinary non-relativistic quantum mechanics and QFT, since we need a theory of (quantum) gravity to predict the behaviour of the black hole, and there are outstanding issues to be resolved about how exactly gravity interacts with quantum mechanics. Certainly the distinction  needs more clarification - in particular, since the point at issue is about whether theoretical expectations about the behaviour of a gravity theory should automatically enter into Alice's predictions even when she has not checked them, there is probably a need to say something more about the relationship between quantum mechanics and gravity before we can determine how this story should go (and conversely, considering how this story \emph{needs} to go in order to maintain consistency could perhaps give us valuable information about how quantum mechanics connects up with gravity). 

This points to a significant problem for the class of `fact-based' relational approaches - specifically, those which are committed to the view that all physical facts are relativized. For if \emph{everything} is relatized, then it becomes more challenging to make a principled distinction between the hypothesis that the in the black hole case the assumptions of Unitarity and EFT should be relativized to the external observer and the hypothesis that in the Wigner's Friend case the fact that Alice has performed a measurement on $S$ should be relativized to Alice. Yet since the relationalist typically wants to preserve the prediction that Bob will find Alice and $S$ to be maximally entangled in the Wigner's Friend case, it appears that they do need to make some distinction. That is, simply relativizing everything seems to deprive us of any means of making principled judgements about what should be relativized and what should not be relativized in any given calculation, which is actually a very important matter since it often leads to substantively different empirical predictions, as we have seen in this case. The more extreme versions of fact-based relationalism may therefore simply be ill-posed and impossible to formulate in any precise way. 
  
Second,  relativizing the Unitarity assumption is in any case not enough by itself to resolve the paradox. To see this, imagine that Alice first prepares the qubit $Q_B$ at time $t_1$, and only at time $t_2$ does she decide whether she will go into the black hole or instead stay outside the black hole and perform the quantum measurements which would verify the assumptions above. So at the earlier time $t_1$, presumably there is some state $\psi$ of the two qubits $Q_B, Q_R$ relative to Alice. Given the  paradigm for black hole physics that Hausmann and Renner have adopted, we know that if Alice goes on to do the other verification measurements she will indeed determine that the assumptions Unitary and EFT hold, so in that case the state $\psi$ will always turn out to be maximally entangled relative to her. Therefore if we say that the relative state at $t_1$ cannot depend on a choice that Alice has not yet made, it follows that the state $\psi$ must always be maximally entangled at $t_1$. Moreover, within the paradigm for black hole physics that Hausmann and Renner have adopted, nothing unusual happens to $Q_B, Q_R$ when Alice enters the black hole, so presumably the state of $Q_B$ and $Q_R$ will still be maximally entangled when  Alice starts doing measurements inside the black hole. This line of argument suggests that even if we allow that the state does not necessarily have to be maximally entangled relative to Alice in the case where Alice does not make appropriate verification measurements, nonetheless if we do not allow her choice to have  a retrocausal influence on the state, then in fact it must  be the case that  $Q_R$ and $Q_B$ are always maximally entangled relative to her.

This suggests that in order to make the kind of solution suggested by \cite{hausmann2025firewallparadoxwignersfriend} work, we will also require something else - probably some kind of retrocausality, because then we can say that if Alice chooses to perform the verification measurements at time $t_2$  the state of $Q_B$ and $Q_R$ at time $t_1$ will be maximally entangled, but if Alice does not choose to perform the verfication measurements at time $t_2$ and instead goes into the black hole, then the state of $Q_B$ and $Q_R$ at time $t_1$ will not be maximally entangled, so Alice never sees a result which is in contradiction with the first-person predictions of unitary quantum mechanics. Alternatively, we might posit something which is not strictly speaking retrocausality but which nonetheless looks that way: for example, one might appeal to ideas associated with quantum gravity to maintain that the relevant connection here in some sense applies prior to the emergence of spacetime and causal structure, so it appears retrocausal from the point of view of the emergent spacetime but in reality involves some underlying nonspatiotemporal structure. But either way, certainly something more than just relationality is required here - so in fact, even if we admit the particularly strong form of relationality suggested by \cite{hausmann2025firewallparadoxwignersfriend}, there is still an important point of difference between paradox 2-BH as compared to paradox 2-WF, since no form of relationality alone will suffice to resolve paradox 2-BH.

A further point in support of the use of retrocausality is   here is the notable fact that this happens to  be a special kind of case in which retrocausality could not be used to create  closed causal loops or paradoxes related to bilking \citep{Ismael2003-ISMCCL-2}. That is, suppose we say that whether or not $Q_B \otimes Q_R$ are maximally entangled at $t_1$ depends on whether  Alice performs a certain measurement at $t_2$. Then under ordinary circumstances, we could imagine that Alice  measures $Q_B$ and $Q_R$ at $t_1$ to see whether they are maximally entangled, determines what the result entails about whether she performs the measurement at $t_2$, and then does the opposite, thus creating a bilking paradox. But in the black hole scenario, if Alice measures $Q_B$ and $Q_R$ at $t_1$ then she has changed the state of $Q_R$ relative to her and so it will not be maximally entangled with $Q_A$ any longer, and thus it is too late for her to go into the black hole and check  if $Q_A$ is entangled with $Q_R$. That is, if Alice measures $Q_B$ and $Q_R$ at $t_1$ then she has in effect chosen that she is going to make the verification measurement rather than going into the black hole to look at $Q_A$, so she will definitely  find them to be maximally entangled. $Q_A$ and $Q_B$ can fail to be maximally entangled only in the case where she does \emph{not} measure them, and hence it seems that mechanism cannot be used to create a closed causal loop or a bilking paradox.  

Moreover, the final state proposal as discussed in section \ref{1BH} furnishes exactly the kind of retrocausal approach that we seem to need here \citep{Bousso_2013}. For in the final state proposal, suppose we take it that the infalling matter system and the Hawking radiation $(Q_B)$ are maximally entangled, and then the interior fields $(Q_A)$ are maximally entangled with the near-horizon fields $(Q_R)$, and meanwhile the final state has the effect of maximally entangling the infalling matter system and the interior fields. Then in the case where Alice performs a  measurement to compare $Q_A$ and $Q_R$, she will find them maximally entangled, and the final state will not have much of an effect; but in the case where she does not perform the measurement to compare $Q_A$ and $Q_R$, in effect we get entanglement swapping and so  $Q_B$ and  $Q_R$ become entangled instead.   So in this picture $Q_R$ is not ever really maximally entangled with two different systems at once, but the way the final state approach works ensures that it is always found to be maximally entangled with the system that we actually compare it to. This account perhaps needs to be supplemented with some story about  how the final state influences or is influenced by the choice made about what to measure, but we can see how this kind of mechanism could plausibly have the right result. And thus the fact that the retrocausal solution seem viable in the black hole case perhaps offers some additional reason to think that we should take retrocausal or teleological solutions seriously in the Wigner's Friend case also. 

 We can also compare paradox 2-BH to other Extended Wigner's Friend paradoxes, such as the theorem of \cite{Bong_2020} which focuses on \emph{first}-person universality. Given that paradox 1-BH actually seems to be a first-person paradox as described above, this comparison might be more apt in any case. In the  Bong et al scenario, in order to maintain first-person universality we must deny one of these assumptions:

\begin{enumerate} 

\item Observers do not branch in the course of performing measurements, and after the measurement the observer sees only one outcome. 

\item Individual observers can accurately communicate their measurement outcomes to one another in an absolute sense. 

\item Locality (of a slightly weaker form than in the standard Bell scenario)

\item No retrocausality. 

\item No superdeterminism. 

\end{enumerate}

So what response is favoured in \emph{this} case if we suppose the solution should be similar in the Wigner's Friend case and the black hole case? Adopting some form of fact-based relationality allows us to deny assumption 2), but we have seen that by itself fact-based relationality doesn't do much to help in the black hole case. And note that in Bousso's version of the black hole scenario all of the measurements involved take place along Alice's worldline, so it seems unlikely  that denying 3) could help there. Thus it seems very likely that, if we  insist on the same kind of solution for both of these paradoxes, we will be pushed toward either a many-worlds solution or a retrocausal or superdeterministic solution for both of these paradoxes. Moreover in section \ref{inherent} I will argue that Paradox 2-BH also provides reasons to be wary about the many-worlds approach, and thus comparing the paradoxes gives us good reasons to think that some form of retrocausality or some kind of teleological picture may be the most natural solution to these paradoxes \footnote{Superdeterminism remains possible, but does not seem a particularly good fit for the kind of case we have been considering  because superdeterminism is by construction quite a non-relational view: in its standard form it rests upon postulating an initial state for the whole universe which encodes all of the desired correlations, and thus it seems to be predicated on the postulation of a global state. Granted, this state is not going to be a \emph{quantum} state, so there is possibly a viable view which combines a global superdeterministic state composed of hidden variables with the stipulation that quantum states specifically are only relational. However, in general if we accept that all quantum states are relational we should perhaps be sceptical about the existence of any kind of global absolute state, for if such a thing existed then one might naturally think it would ground a global quantum state as well. So although there is perhaps not a strict incompatibility between dynamical relationalism and superdeterminism, there is certainly some kind of conceptual tension.}.

\subsubsection{Inherent Relationality \label{inherent}}

There is also another way of presenting the paradox 2-BH: within the paradigm for black hole physics that Hausmann and Renner have adopted, it follows immediately that after the protocol describved above the qubit $Q_R$ is maximally entangled with qubit $Q_A$ and \emph{also} maximally entangled with qubit $Q_B$, so we seem to arrive at a state $\psi$ in which $Q_R$ is maximally entangled with two different systems at once. But the usual quantum formalism has a feature known as `monogamy of entanglement,' meaning that a given  system cannot be maximally entangled with more than one system, so this seems to contradict the predictions of standard quantum mechanics: this state $\psi$ simply cannot be a valid quantum state. 

Now, we can certainly resolve this problem if we posit  inherent relationality. For the  putative maximally entangled state $\psi$ of $Q_R, Q_B$ and $Q_A$ cannot be the state relative to any actual observer: it may be true relative to an external observer who has verified the properties of the asymptotic radiation that $Q_R$ and $Q_B$ are maximally entangled, and simultaneously it may be true relative to Alice that $Q_A$ and $Q_R$ are maximally entangled, but if we follow the suggestion of Hausmann and Renner and posit that only observers who can verify the Unitarity and EFT assumptions will see $Q_R$ and $Q_B$ as maximally entangled, then there is nobody relative to whom both relations of maximal entanglement hold at the same time. Thus if  quantum states are inherently relational, then since the state $\psi$ cannot be relativized to any observer, there \emph{is} no such state and therefore it cannot violate the monogamy of entanglement.  

By contrast, an approach which posits only effective relationality, such as the Everett approach or de Broglie-Bohm, is still committed to the existence of an underlying absolute quantum state from which all of the individual relational states are derived. And one might naturally expect that this underlying absolute state will say  that $Q_R$ is maximally entangled with both $Q_A$ and $Q_B$, thus explaining why $Q_A$ and $Q_R$ and separately $Q_R$ and $Q_B$ are both seen as maximally entangled by different observers. But then the underlying absolute state would violate the monogamy of entanglement and thus is not a valid quantum state. This indicates immediately that  inherently relational approaches have an advantage over the effective relational approaches with regard to this specific problem, because we have just seen adopting inherent relationality is by itself enough to resolve this problem, whereas adopting effective relationality by itself does little to resolve the problem. 

Of course, it remains possible that there is some other way to resolve the problem within the effective approaches, but certainly it is not automatic and will require some further work. For example, in the Everettian case  a natural first response is to note that of course in the Everett picture there is no problem with allowing that   a given system can be maximally entangled with two distinct systems \emph{within different branches of the wavefunction}. Suppose that I perform a measurement on a spin qubit, and then if the result is `up' I perform an operation which results in particles $A$ and $B$ becoming maximally entangled, whereas if the result is `down' I instead perform an operation which results in particles $A$ and $C$ becoming maximally entangled. This is not a violation of the monogamy of entanglement because the entanglement exists in a branch-relative way and $A$ is never maximally entangled with two different systems within a single branch. So similarly, the Everettian might hope to argue in the black hole case that  $Q_A - Q_R$ are entangled relative to one branch and  $Q_B - Q_R$ are entangled relative to another branch and thus we never have violations of monogamy relative to any one branch.

However, this solution does not seem to be available given the setup of the paradox put forward by Hausmann and Renner. This is because it is assumed as part of the setup that all of the observers can still communicate with each other throughout the experiment - Bob can send a record to Alice and she can read it. And in the Everettian picture, observers in different branches cannot communicate with each other, so we must conclude that there is at least one branch in which Alice and Bob both exist so that they can communicate with each other. But then we  can just focus on that branch and make all our arguments just relative to that branch. The same arguments still seem to go through - within any one Everettian branch observers are supposed to see predictions consistent with standard quantum mechanics, so we'd expect that within that branch observers will see   $Q_A - Q_R$ as maximally entangled and also see $Q_B - Q_R$ as maximally entangled, so the problem recurs. Of course, without a complete theory of quantum gravity and full story about how the Everett approach interacts with it, we can't completely rule out the possibility that gravitational effects interact with Everettian branching in a way that somehow permits us to violate the prediction that observers will see  $Q_A - Q_R$ and $Q_B - Q_R$, but this is by no means automatic. Thus we see clearly that insofar as our concern is to solve the monogamy problem, inherent relationality offers an immediate resolution whereas effective relationality by itself does not resolve the issue. 
  
This is interesting because in many of the original Wigner's Friend paradoxes, such as  1-WF and 2-WF and the theorem of \cite{Bong_2020} it appears that either effective or inherent relationality will suffice to resolve the paradox - this is clear from the fact that both the Everettian and de Broglie-Bohm approaches have a perfectly consistent story to tell about these experiments, despite positing only effective relationality. But the black hole version looks different: in that case the monogamy problem is immediately resolved by positing inherent relationality, but effective relationality in and of itself does not really help at all.    Therefore if we suppose that these two similar cases 2-WF and 2-BH should be explained in similar ways, there is at least some reason to think we might be better off  adressing the Wigner's Friend paradoxes by appeal to inherent relationality rather than merely effective relationality.

\section{Conclusion}

In this article I have presented a further analysis of the analogy between  the Wigner's Friend paradoxes and black hole paradoxes originally pointed out by \cite{hausmann2025firewallparadoxwignersfriend}, arguing that   comparing the original Wigner's Friend paradoxes with black hole cases can help us decide how we ought to resolve the Wigner's Friend cases. Now of course none of the arguments herein are intended to be conclusive: it could be that the similarity is only superficial and in fact different mechanisms are involved in the two cases. That would not be unreasonable, since the black hole case involves gravitational phenomena and the Wigner's Friend case does not. But nonetheless, since very similar issues seem to be at stake, it is at least plausible to think that the solutions should be similar. 

We have seen in this article that if it is assumed  that the solutions should look similar, we learn two main things. First, that there are good reasons to resolve these paradoxes using an \emph{inherently} relational approach, rather than an effective one like the Everettian or de Broglie-Bohm picture. And second, that there are good reasons to adopt a \emph{retrocausal} or teleological approach. 

In particular, note that a number of recent presentations of the Bong et all EWF paradox have suggested that we should resolve it by giving up `absoluteness' i.e. by adopting what I have called fact-based relationality. However, as argued by \cite{adlam2025kindrelationalitydoesquantum} there are strong epistemic reasons to be wary about fact-based relationalism. And we have seen here that fact-based relationalism will not suffice to resolve the paradoxes in the black hole case anyway - we are still going to need something at at least superficially looks like retrocausality or some kind of teleological effect. This suggests strongly that in the original EWF cases, the right way forward is not to posit fact-based relationality but rather to stick with some kind of dynamical relationality and deny the no-retrocausality assumption. That is,  one might be tempted to take this as a sign that fact-based relationality was always a dead end and we are better off focusing on how to make dynamical relationality work in combination with appropriate retrocausal or teleological effects. 

Note also that in addition to gaining new insight into the Wigner's Friend scenarios from this comparison, we have also potentially learned something about black holes. In particular, we have noted that in the Wigner's Friend scenarios what we need to resolve the paradox is simply \emph{dynamical} relationalism, and that it is plausible that dynamical relationalism may suffice in the black hole case as well. That is, rather than saying that there does not exist any joint description at all across the interior and exterior of a black hole, we need only say that there does not exist any global \emph{quantum state} which correctly describes both regions. Thus the comparison potentially sheds some light on ongoing discussions of black hole complementarity - depending on how one defines the word `complementarity,' it seems that either dynamical relationalism offers a new way of thinking about complementarity, or else simply a different kind of relational approach which may serve as an alternative to complementarity.  This demonstrates that the black hole paradoxes do not necessarily have to be thought of as pointing us toward some kind of radical metaphysics - relativizing some dynamics while leaving the underlying facts non-relational could be enough. 

\section{Acknowledgements:} 

Many thanks to Ladina Hausmann for very helpful comments on a draft of this paper. This work was supported by the  John Templeton Foundation Grant ID 63209.

\end{document}